\documentclass[lettersize,journal]{IEEEtran}
\usepackage{amsmath,amsfonts}
\usepackage{algorithmic}
\usepackage{algorithm}
\usepackage{array}
\usepackage[caption=false,font=normalsize,labelfont=sf,textfont=sf]{subfig}
\usepackage{textcomp}
\usepackage{stfloats}
\usepackage{url}
\usepackage{verbatim}
\usepackage{graphicx}
\usepackage{cite}
\usepackage{diagbox}
\usepackage{amssymb}
\usepackage{enumitem}
\begin{document}

\title{Parallel Multi-Extended State Observers based {ADRC} with Application to High-Speed Precision Motion Stage}

\author{Guojie Tang, ~\IEEEmembership{Student Membership,~IEEE,}
	Wenchao Xue, ~\IEEEmembership{Membership,~IEEE,}
	Hao Peng, \\
	 Yanlong Zhao, ~\IEEEmembership{Membership,~IEEE,}
	and Zhijun Yang, ~\IEEEmembership {Membership,~IEEE}

\thanks{Guojie Tang, Wenchao Xue and Yanlong Zhao are with the Key Laboratory of Systems and Control, Academy of Mathematics and Systems Science, Chinese Academy of Sciences, Beijing 100190, P. R. China and the School of Mathematical Sciences, University of Chinese Academy of Sciences, Beijing {\rm 100049}, China (e-mail: tangguojie18@mails.ucas.ac.cn; wenchaoxue@amss.ac.cn; ylzhao@amss.ac.cn).

Hao Peng and Zhijun Yang  are with the School of Electromechanical Engineering, Guangdong University of Technology Guangzhou, Guangdong 510006, China (e-mail:penghao@mail2.gdut.edu.cn; yangzj@gdut.edu.cn).

}

}

\markboth{}%
{Shell \MakeLowercase{\textit{et al.}}: A Sample Article Using IEEEtran.cls for IEEE Journals}


\maketitle

\begin{abstract}
In this paper, the parallel multi-extended state observers (ESOs) based active disturbance rejection control approach is proposed to achieve desired tracking performance by automatically selecting the estimation values leading to the least tracking error. First, the relationship between the estimation error of ESO and the tracking error of output is quantitatively studied for single ESO with general order. In particular, the algorithm for calculating the tracking error caused by single ESO's estimation error is constructed. Moreover,  by timely evaluating the least tracking error caused by different ESOs, a novel switching ADRC approach with parallel multi-ESOs is proposed. In addition, the stability of the algorithm is rigorously proved. Furthermore, the proposed ADRC is applied to the high-speed precision motion stage which has large nonlinear uncertainties and elastic deformation disturbances near the dead zone of friction. The experimental results show that the parallel multi-ESOs based ADRC has higher tracking performance than the traditional single ESO based ADRC.
\end{abstract}


\section{Introduction}
\IEEEPARstart{U}{ncertainties}, including unmodeled dynamics, time-varying parameters and external disturbances, widely exist in various industrial control systems.  How to force the system to track the desired output signal under the influence of uncertainties is a basic problem throughout the development of control science. To cover the uncertainties in system dynamics, a number of control methods were presented, such as proportional-integral-derivative (PID) control \cite{inproceedings1}, adaptive control \cite{book1}, sliding mode control \cite{book2}, disturbance observer basic control (DOBC) \cite{book3} and active disturbance rejection control (ADRC) \cite{article1, inproceedings2}. Among them, since ADRC can estimate and compensate the ``total disturbance'' including internal uncertain dynamics and external disturbances in real time through its core module extended state observer (ESO),  it has received wide attention. Apart from theoretical studies about the stability and convergence of ADRC \cite{inproceedings3, article2}, a significant increase in applications of ADRC across different fields can be noticed in recent years, including but not limited to spacecraft control systems \cite{article3,inproceedings4, article24}, power electronics \cite{article4,article5}, motion control systems \cite{article6, article7, article16,article21} and others \cite{article8,article22}.


Since the key idea of ADRC is timely estimating both states and total disturbance, how to improve ESO to achieve better estimation is an important topic. One kind of representative work is to improve the observation accuracy of ESO by designing time-varying observer gains: \cite{article9} proposed the adaptive ESO (AESO) based ADRC, in which the gain of ESO was timely tuned to reduce the estimation errors of both states and the ``total disturbance" against the measurement noise, and applied it to the air-fuel ratio (AFR) control of gasoline engine. \cite{article17} designed ESO with predetermined decreasing gains to reduce the influence of measurement noise in system steady state and the method was tested in a magnetic levitation ball system.  \cite{article18} proposed an adaptive ADRC parameters adjustment algorithm based on off-line Q-learning and applied it to the ship course control. Moreover, there were analogous attempts in \cite{article19, article20}.

In addition to studying the most typical ESO, which is usually one order higher than the system order, some scholars have also focused their attention on high-order ESO, which can estimate the high-order derivative of the total disturbance.  By analyzing and comparing the frequency regional performance of different order ESOs, \cite{article16}  drew the conclusion that as the order of ESO increased, its disturbance rejection ability also increased, but the estimation performance also became more sensitive to noise. \cite{article8}, \cite{article12} and  \cite{article13} studied the stability of high-order ESOs in time-domain and pointed out the influence of each parameter on ESO and the closed-loop system.

 Moreove, the structure of ESO has also been extensively studied. The most typical examples are the non-linear ESO (NLESO)  and the linear ESO (LESO) \cite{article1}.
\cite{article10} proposed a linear/non-linear switching extended state observer (L/NL-SESO) to solve the problem that NLESO was limited to large amplitude disturbance estimation and the performance of the corresponding ADRC algorithm was improved indirectly. Aiming to reduce the impact of measurement noise on ESO, \cite{article11} proposed the cascade ESO (CESO) in which a unique cascade combination of ESOs was developed and the estimation of the `` total disturbance'' was the sum of the ESOs' estimations. \cite{inproceedings5} studied the time-domain responses of the 3rd-order LESO and the 4th-order LESO to step disturbance and proposed a switching ADRC based on parallel 3rd-order LESO and 4th-order LESO. As for the defects of traditional ESO in estimating large scale fast-varying sinusoidal disturbance (FVSD), \cite{article23} came up with the generalized integrator-extended state observer (GI-ESO), which enabled FVSD to be observed with a relatively low bandwidth, and applied it to the Grid-connected Converters.

However, as mentioned above, almost all of these improvements indirectly improved the control effect of ADRC by improving the estimation ability of ESO, rather than directly optimizing the control law. Along this line, two problems arise: How does the estimation effect of ESO affect the control performance of ADRC, in other words, does a smaller estimation error mean a smaller tracking error? If the answer is negative, then how to optimize the control law directly?

In this paper, we explore the above two questions and propose the parallel multi-ESOs based ADRC design approach. Our contribution can be summed up as follows:

(i) The relationship between the estimation error of ESO and the tracking error of controlled output is quantitatively studied for single ESO with general order. In particular, the algorithm for calculating the deviation of tracking caused by ESO's estimation error is proposed.
%

(ii) By timely evaluating the least tracking error caused by different ESOs, a novel switching ADRC approach with parallel multi-ESOs is proposed. Furthermore, the stability of the algorithm is rigorously proved under some standard assumptions.


(iii) The proposed control scheme is implemented in a high-speed precision motion stage for higher precision of tracking.. The experimental results show the validity and strong robustness of our method, with the parallel multi-ESOs based ADRC has higher control performance.

%


The rest of this paper is arranged as follows: Section II presents the basic problem formulation to study. Section III analyzes the influence of the estimation error of LESO on the control performance and gives the designing approach of the parallel multi-ESOs based ADRC.
Experimental studies are shown in Section IV and some conclusions are given
in Section V.

\section{Problem Formulation}
Considering the following SISO system combined with uncertain dynamics and external disturbances:
\begin{equation}\label{system}
\left\{ \begin{array}{l}
\dot x(t) = Ax(t) + B(bu(t) + f(x(t),t)\\
y(t) = Cx(t)
\end{array}, \right.
\end{equation}
where
\begin{small}
\begin{equation}
 A= {\left[ {\begin{array}{*{20}{c}}
0&1&0& \cdots &0\\
0&0&1& \cdots &0\\
 \vdots & \vdots & \vdots & \vdots & \vdots \\
0&0&0& \cdots &1\\
0&0&0& \cdots &0
\end{array}} \right]_{n \times n}},B = {\left[ {\begin{array}{*{20}{c}}
0\\
0\\
 \vdots \\
0\\
1
\end{array}} \right]_{n \times 1}},C = {\left[ {\begin{array}{*{20}{c}}
1\\
0\\
 \vdots \\
0\\
0
\end{array}} \right]^T_{n \times 1}}.
\end{equation}
\end{small}
$t$ is time, $x=[x_1, x_2, ..., x_n] \in R^n$ is the state vector, $u \in R$ and $y \in R$ are the input and output of the system, respectively. $b \in  R$ is the known input gain. $f \in R$  represents internal uncertain dynamics and external disturbances. 

This paper considers the control problem in which a signal $r_1(t)$ is taken as the
reference command for the system output $y(t)=x_1(t).$ In practice, the output $y=x_1$ is usually expected to track the desired reference signal despite various uncertainties. Usually, $y(t)$ is requested to achieve desired both transient and steady performances which can be described by the following ideal trajectory:

\begin{equation}\label{ideal system}
\left\{ \begin{array}{l}
{{\dot x}^*} = A{x^*} + B\left( { - {K^T}({x^*} - r) + {r_{n + 1}}} \right)\\
{x^*}({t_0}) = x({t_0})
\end{array} \right..
\end{equation}
$r=[r_1,r_2,..,r_n]$, $r_i={r_1}^{(i-1)}$, $K^T=[k_1,k_2,...,k_n]$ and the matrix $(A-BK^T)$ is a Hurwitz matrix whose characteristic polynomial has following form:

\begin{equation}\label{Characteristic polynomial}
\Delta (s) = (s+s_1)^{d_1}(s+s_2)^{d_2}\cdots (s+s_l)^{d_l},
\end{equation}
where $l>0$,  $Re[s_i]>0, i=1,2,..,l$ and $d_1 + d_2 +\cdots +  d_l =n$. Apparently, the performance of $x^*$ can be optimized by tuning $K^T$.

 In this paper, we focus on the following control objective:
\begin{equation}\label{objective 1}
\mathop {\sup }\limits_{t \in [{t_0},\infty )} |{x_1}(t) - x_1^*(t)| \leq \eta ,
\end{equation}
where $\eta > 0$ is the maximum acceptable tracking error between $x_1$ and its ideal trajectory $x_1^*$. 
 
Defining $e=[e_1,...,e_n]^T =x-r$ and $e^*=[e_1^*,...,e_n^*]^T=x^*-r$, we have $e-e^*=x-x^*$. Therefore, the control objective (\ref{objective 1}) can be equivalent to the tracking problem of $e_1$ to $e_1^*$ :
\begin{equation}\label{objective 2}
\mathop {\sup }\limits_{t \in [{t_0},\infty )} |{e_1}(t) - e_1^*(t)| \leq \eta .
\end{equation}

Now, the problem is how to design control law $u$ to achieve control object (\ref{objective 2}). First, according to (\ref{system}) and (\ref{ideal system}), the ideal closed-loop error system could be written as :
\begin{equation}\label{error system 1}
{{\dot e}^*} = (A - B{K^T})e^*,\quad e^*(t_0)=e(t_0).
\end{equation}
Then, we rewrite (\ref{system}) in terms of the tracking error $e$ :

\begin{equation}\label{error system 2}
\dot e = Ae + B(bu + e_{n+1}),
\end{equation}
where $e_{n+1} = f-r_{n+1}.$
Obviously, if $e$ and $e_{n+1}$ are available, the following $u^*$ will make (\ref{error system 2}) be the same to (\ref{error system 1}) :

\begin{equation}
{u^*}(t) = \frac{{ - {K^T}e(t) - e_{n+1}(t)}}{b}.
\end{equation}
However, it is difficult to obtain the high-order differential of the reference signal $r_1$ and the system states $x_i,i=2,3,...,n$ can not be measured, which both lead to that each differential of the tracking error $e_1$ can't be obtained. Besides that, the total disturbance $e_{n+1}$ is also unknown. Despite these difficulties, we could estimate $e(t)$ and $e_{n+1}(t)$ in real time by designing the LESO and design ADRC law to achieve the control object.

\section{Parallel Multi-ESOs based ADRC Design}

\subsection{Single LESO based ADRC Design}

In this section, a $(n+m)$-th order LESO based ADRC is designed to reach the control object. It is worth noting that the following study always holds for ADRC based on $(n+i)$th order LESO, $i=1,2,...,m$.

 Generally, a $(n+m)$th order LESO has the following form :
\begin{equation}\label{LESO}
\begin{cases}
\dot{\hat e}(t) =  ({A_{n + m}} - \beta {C_{n + m}})\hat e(t) \\
 \qquad \quad + \beta {C_{n + m}}{e}(t) + {B_{n + m}}bu(t)\\
\hat e_1({t_0})  =  {e_1}({t_0})\\
\hat e_i({t_0})  =  {e_{i,0}},\quad i = 2,3,...,n + m
\end{cases},
\end{equation}
where $e_{i,0}$ is the estimation of $e_i(t_0)$,
\begin{small}
\begin{equation}
\begin{array}{*{20}{l}}
{{A_{n + m}} = {{\left[ {\begin{array}{*{20}{c}}
0&1&0& \cdots &0\\
0&0&1& \cdots &0\\
 \vdots & \vdots & \vdots & \vdots & \vdots \\
0&0&0& \cdots &1\\
0&0&0&0&0
\end{array}} \right]}_{(n + m) \times (n + m)}}}\\
{{B_{n + m}} = \left[ {\begin{array}{*{20}{c}}
B\\
{{0_{m \times 1}}}
\end{array}} \right],{C_{n + m}} = {{\left[ {\begin{array}{*{20}{c}}
{{C^T}}\\
{{0_{m \times 1}}}
\end{array}} \right]}^T},\beta  = \left[ {\begin{array}{*{20}{c}}
{{\beta _1}}\\
 \vdots \\
{{\beta _{n + m}}}
\end{array}} \right]}
\end{array},
\end{equation}
\end{small}
and
\begin{equation}
{{{(s + {\omega _o})}^{n+m}} = {s^{n+m}} + {\beta _1}{s^{n + m - 1}} +  \cdots  + {\beta _{n + m}}}.
\end{equation}

The observer bandwidth $\omega_o > 0$  is the parameter to be adjusted and $\hat e =[\hat e_1,..., \hat e_{n+m}]$ is the estimation value of $[e_1,...,{e_{n+m}}]$, where $e_{n+i}=e_{n+1}^{(i-1)},i=1,...,m$.

Then, the following traditional signal ESO based ADRC law could be designed as (\ref{ADRC}) 
\begin{equation}\label{ADRC}
u(t) = \dfrac{{ - {k_1}{{\hat e}_1}(t) -  \cdots  - {k_n}{{\hat e}_n}(t) - {{\hat e}_{n + 1}}(t)}}{b}.
\end{equation}
and the error system between (\ref{error system 1}) and (\ref{error system 2}) could be described as (\ref{error system}) :
\begin{equation}\label{error system}
\dot{\bar{e}} = {A^*}\bar{e} + {B}\delta(\tilde e),
\end{equation}
where $\bar e=e-e^*$, $A^*=A-BK^T$ and $\delta (\tilde e) = \sum\limits_{i = 1}^n {{k_i}{{\tilde e}_i} + {{\tilde e}_{n + 1}}} $.

\begin{figure}[H]\centering
\begin{minipage}[t]{\linewidth}
 \includegraphics[width=1\linewidth]{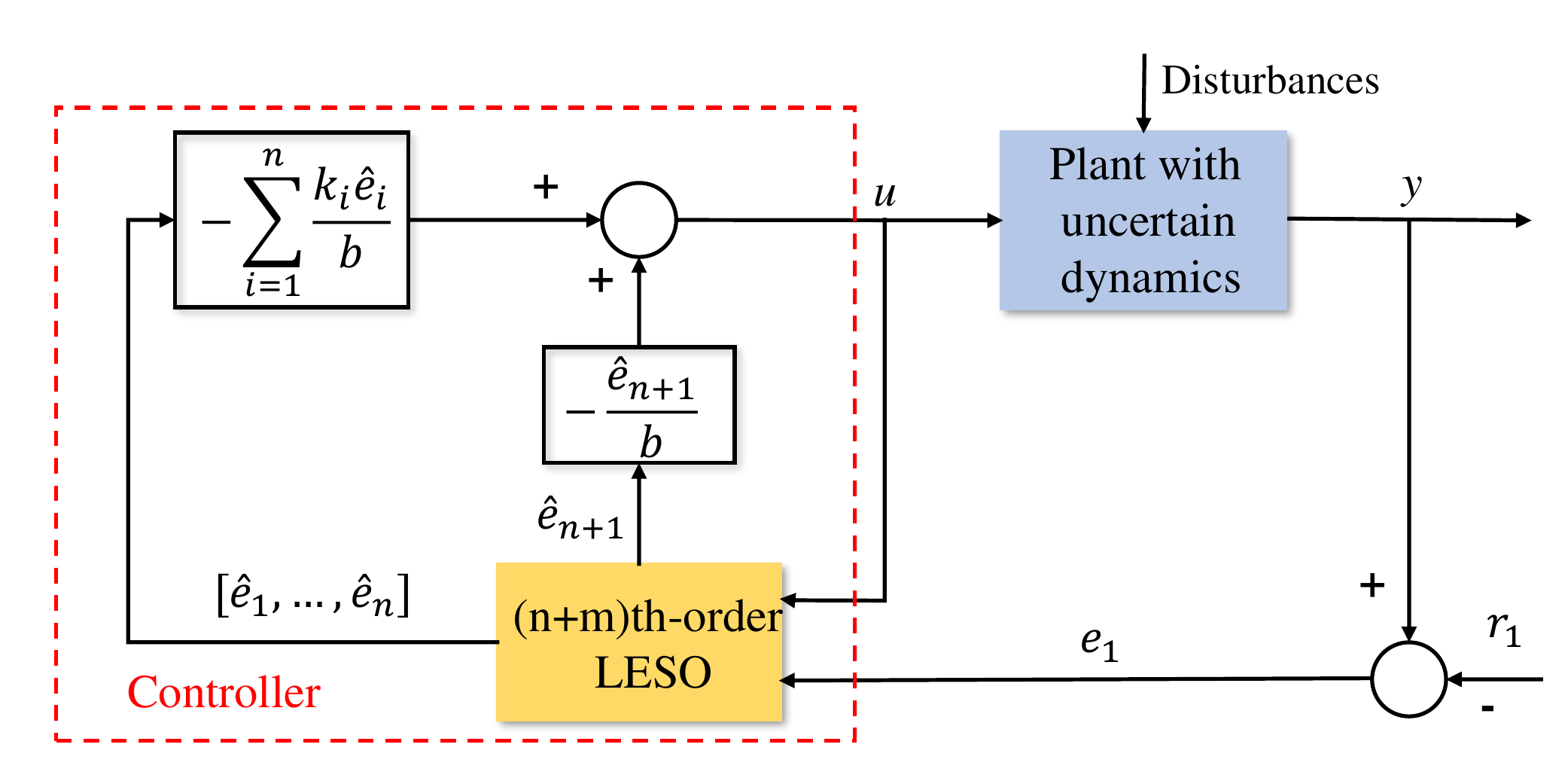}
      \caption{Control block diagram of traditional signal ESO based ADRC  }
      \label{control block disgram 1}
\end{minipage}
\end{figure}

The control block diagram is shown in Fig.~\ref{control block disgram 1}.

For the convenience of subsequent analysis, we redefine the tracking error vector $e=[e_1,...,e_{n+m}]$. Let $\tilde e = e - \hat e$ be the estimation error vector, then the dynamic of LESO's estimation error could be described as follow:

\begin{equation}\label{estimation error}
\dot {\tilde e} = ({A_{n + m}} - {\beta }{C_{n + m}})\tilde e + \tilde B{e_{n+1}^{(m )}},
\end{equation}
where $\tilde {B}=[0,\cdots,0,1]^T_{(n+m) \times 1}$.

Actually, to guarantee the stability of ADRC, it is essential to prove the boundedness of the LESO's estimation error. 

Define
\begin{equation}
{\varepsilon _i} = \frac{{{{\tilde e}_i}}}{{{\omega _o}^{i - 1}}},\quad \varepsilon  = [{\varepsilon _1}, \cdots ,{\varepsilon _{n + m}}]^T.
\end{equation}
Then, the estimation error equation (\ref{estimation error}) could be rewritten as
\begin{equation}\label{eatimation error Re}
\dot \varepsilon  = {\omega _o}\tilde A\varepsilon  + \tilde B\frac{{{e_{n+1}^{(m)}}}}{{\omega _o^{n+m-1}}},
\end{equation}
where 
\begin{equation}
{\tilde A = \left[ {\begin{array}{*{20}{c}}
{ - {\alpha _1}}&1&0& \cdots &0\\
{ - {\alpha _1}}&0&1& \cdots &0\\
 \vdots & \vdots & \vdots & \ddots & \vdots \\
{ - {\alpha _{n + m - 1}}}&0&0& \cdots &1\\
{ - {\alpha _{n + m}}}&0&0& \cdots &0
\end{array}} \right]}
\end{equation}
and
\begin{equation}
{{\alpha _i} = \frac{{{\beta _i}}}{{\omega _o^i}},i = 1,2,...,n + m}.
\end{equation}

In fact, the stability of the designed $(n+m)$-th order LESO has been well studied in \cite{article12} under the following Assumptions 1-2:

{\bf Assumption 1.} There is a positive integer $m$ such that the first $m$th derivatives of $f$ are bounded, i.e., there exists  $h_1> 0$  such that
\begin{equation}
\mathop {\sup }\limits_{t \in [{t_0},\infty )} |{f^{(i)}}| \leq {h_1},\quad, i=1,2,...,m;
\end{equation}

{\bf Assumption 2.} The exists a known positive $h_2$ satisfying
\begin{equation}
\mathop {\sup }\limits_{t \in [{t_0},\infty )} \left| {{r_i}(t)} \right| \leq h_2,\quad i = 1,2,...,n + m.
\end{equation}

 Here, we quote their conclusion as Lemma 1 directly.

{\bf Lemma 1.} (\cite{article12}) Considering estimation error systems (\ref{estimation error}) and (\ref{eatimation error Re}) under Assumption 1-2, we have the following results for any $t>t_0$, $1 \leq i \leq n+m$:\\
(i)\begin{small}
\begin{equation}
\vert \tilde e_i(t)  \vert \leq {\omega_o}^{i-1} \Vert e^{\omega_o\tilde{A}(t-t_0)}\Vert_{\infty}\Vert \varepsilon(t_0) \Vert_{\infty} + \dfrac{(h_1 + h_2 )G_i}{{\omega_o}^{n+m-i+1}},
\end{equation}
\end{small}
where ${G_i} = \sum\limits_{j = 0}^{i - 1} {\left( {\begin{array}{*{20}{c}}
{n + m - i + j}\\
{n + m - i}
\end{array}} \right)} $;\\
(ii)
\begin{equation}
{\Vert {\varepsilon (t)} \Vert_\infty } \leq {\Vert {{e^{{\omega _o}\tilde A(t - {t_0})}}} \Vert_\infty }{\Vert {\varepsilon ({t_0})} \Vert_\infty } + \frac{{(h_1 + h_2 )G}}{{\omega _o^{n + m}}},
\end{equation}
where$\quad G = \mathop {\max }\limits_{1 \leq i \leq n + m} {G_i}$.

\subsection{Parallel Multi-ESOs based ADRC Design}

In order to give full play to the advantages of LESOs with  different orders and parameters, a natural idea is to set parallel multi-LESOs and choose different LESO based ADRC laws according to the real time data. Therefore, it is necessary to analyze how the different orders and parameters of LESO will affect the control effect.

In the rest of this article, we use $L\left[ \cdot \right]$ and  $L^{-1}\left[ \cdot \right]$ to represent the Laplace transform and inverse Laplace transform, respectively. Lemma 2 below describes the influence of LESO's estimation error on the control error quantitatively.

{\bf Lemma 2.} Considering the estimation error systems (\ref{estimation error})  and the closed-loop error system (\ref{error system}) under the same condition of Lemma 1, the tracking error $\bar e_1$ satisfies the following formula:
 \begin{small}
 \begin{equation}\label{e1 inv laplace}
{\bar e_1}(t) = {L^{ - 1}}\left[\dfrac{{{g_n}(s)}}{{\Delta (s)}}{\tilde E_1}(s)\right] + \sum\limits_{i = 2}^n {{L^{ - 1}}\left[\dfrac{{{g_{n - i}}(s)}}{{\Delta (s)}}{e^{ - {t_0}s}}\right]} \tilde e_i({t_0})
\end{equation}
\end{small}  
where
\begin{equation}
{g_n}(s) = \sum\limits_{i = 0}^n {\left( {\sum\limits_{j = 0}^i {{\beta _j}{k_{n + 1 - i + j}}} } \right)} {s^{n - i}}, \beta_0 = k_{n+1} =1,
\end{equation}
and 
\begin{equation}
\begin{array}{*{20}{c}}
{{g_i}(s) = \sum\limits_{j = 0}^i {{k_{n + 1 - j}}{s^j}} ,\quad i = 0,1,...,n - 1}
\end{array}.
\end{equation}

{\bf Remark 1.} Lemma 2 demonstrates the relation between $\bar e_1$ and $\tilde e_1$, as well as the initial values. In addition, we will prove the boundedness of $\tilde e$ and $\bar e$ later, which guarantees the existence of the Laplace transform.

{\bf Proof of Lemma 2.}

Just considering the first $n$ equations of (\ref{estimation error}) and taking Laplace transform to them, then we have:
\begin{equation}
\left\{ {\begin{array}{*{20}{l}}
{s{{\tilde E}_1}(s) - {e^{ - {t_0}s}}{{\tilde E}_1}(t_0) = {{\tilde E}_2}(s) - {\beta _1}{{\tilde E}_1}(s)}\\
 \vdots \\
{s{{\tilde E}_n}(s) - {e^{ - {t_0}s}}{{\tilde E}_n}(t_0) = {{\tilde E}_{n + 1}}(s) - {\beta _n}{{\tilde E}_1}(s)}
\end{array}.} \right.
\end{equation}
After sorting it out, $\tilde E_{i+1}(s), i = 1,2,...,n$ can be expressed by $\tilde E_1(s)$ and the initial estimation error of each state:
\begin{equation}
{\tilde E_{i + 1}}(s) = (\sum\limits_{j = 0}^{i} {{\beta _j}{s^{i - j}}} ){\tilde E_1}(s) - {e^{ - {t_0}s}}\sum\limits_{j = 1}^{i} {{s^{i - j}}{{\tilde E}_j}(t_0)}.
\end{equation}

In the same way, taking Laplace transform to (\ref{error system}) and then we have
\begin{footnotesize}
\begin{equation}
\left\{ {\begin{array}{*{20}{l}}
{s{{\bar E}_1}(s) - {e^{ - {t_0}s}}{{\bar E}_1}({t_0}) = {{\bar E}_2}(s)}\\
 \vdots \\
{s{{\bar E}_{n - 1}}(s) - {e^{ - {t_0}s}}{{\bar E}_{n - 1}}({t_0}) = {{\bar E}_n}(s)}\\
{s{{\bar E}_n}(s) - {e^{ - {t_0}s}}{{\bar E}_n}({t_0}) =  - \sum\limits_{i = 1}^n {{k_i}{{\bar E}_i}(s) + \sum\limits_{i = 1}^{n + 1} {{k_i}{{\tilde E}_i}(s)} } }
\end{array}.} \right.
\end{equation}
\end{footnotesize}

After sorting out the above equations, $E_1(s)$ expressed by $\hat E_1(s)$ and the initial estimation error of each state is obtained:
\begin{footnotesize}
\begin{equation}\label{laplace e1}
\begin{aligned}
{\bar E_1}(s) &= \dfrac{{{g_n}(s)}}{{\Delta (s)}}{{\tilde E}_1}(s) + {e^{ - {t_0}s}}\sum\limits_{i = 1}^n {\dfrac{{{g_{n - i}}(s)}}{{\Delta (s)}}({{\bar E}_i}({t_0}) - \tilde E({t_0}))} \\
 &= \dfrac{{{g_n}(s)}}{{\Delta (s)}}{{\tilde E}_1}(s) + {e^{ - {t_0}s}}\sum\limits_{i = 2}^n {\dfrac{{{g_{n - i}}(s)}}{{\Delta (s)}}\tilde e_i({t_0})} 
\end{aligned}.
\end{equation}
\end{footnotesize}

Taking Inverse Laplace transform to (\ref{laplace e1}), the result is obtained. $\blacksquare$

However, by (\ref{e1 inv laplace}),  it needs to know the unmeasurable $\tilde{e}_i(t_0)$ to compute $e_1$. Ignoring the influences of these initial values, an approximation of $e_1$ can be indicated as 

\begin{equation}\label{approximate}
z(t) = {L^{ - 1}}\left[\dfrac{{{g_n}(s)}}{{\Delta (s)}}{\tilde E_1}(s)\right] .
\end{equation}
Although there is an approximate error, the following Theorem 1 demonstrates that $z$ will converge to $\bar e_1$ as time goes on.

{\bf Theorem 1.} Considering the calculation formula (\ref{e1 inv laplace}) of tracking error $\bar e_1$ and its approximation formula (\ref{approximate}), there are 

(i) \begin{equation}
z(t) - {\bar e_1}(t)   = \sum\limits_{i = 2}^{n - 1}\sum\limits_{j=1}^l e^{-s_j(t-t_0)} p_{i,j}(t-t_0) \tilde{e}_i(t_0) ,
\end{equation}
where
\begin{equation}
{p_{i,j}}(t - {t_0}) = \sum\limits_{k = 1}^{{d_j}} {\frac{{{c_{i,j,{d_j} - k + 1}}}}{{\left( {{d_j} - k} \right)!}}{{(t - {t_0})}^{{d_j} - k}}}.
\end{equation}

(ii)
\begin{equation}
\lim _{t \to \infty} \vert z(t) - {\bar e_1}(t)  \vert =0.
\end{equation}
{\bf Remark 2.}  Theorem 2 illustrates that ${z}$ will converge to the true value $\bar e_1$ over time, and the convergence speed can be accelerated by rational pole allocation. Thus, $\bar e_1$ can be calculated approximately by only using the available $\tilde e_1$.

{\bf Proof of Theorem 1.}

Decomposing $g_{n-i}(s)/\Delta (s)$ and the following result could be obtained:

\begin{equation}
\dfrac{{{g_{n - i}}(s)}}{{\Delta (s)}} = \sum\limits_{j = 1}^l {\sum\limits_{k = 0}^{{d_j} - 1} {\dfrac{{{c_{i,j,{d_j} - k}}}}{{{{(s + {s_j})}^{{d_j} - k}}}}} } 
\end{equation}

where
\begin{footnotesize}
\begin{equation}\label{c}
\left\{ {\begin{array}{*{20}{l}}
{{c_{i,j,{d_j}}} = \mathop {\lim }\limits_{s \to  - {s_j}} {{(s + {s_j})}^{{d_j}}}\dfrac{{{g_{n - i}}(s)}}{{\Delta (s)}}}\\
{{c_{i,j,{d_j} - 1}} = \mathop {\lim }\limits_{s \to  - {s_j}} \dfrac{\mathrm{d}}{\mathrm{ds}}[{{(s + {s_j})}^{{d_j}}}\dfrac{{{g_{n - i}}(s)}}{{\Delta (s)}}]}\\
 \vdots \\
{{c_{i,j,1}} = \frac{1}{{({d_j} - 1)!}}\mathop {\lim }\limits_{s \to  - {s_j}} \dfrac{{\mathrm{d^{({d_j} - 1)}}}}{\mathrm{d{s^{({d_j} - 1)}}}}[{{(s + {s_j})}^{{d_j}}}\dfrac{{{g_{n - i}}(s)}}{{\Delta (s)}}]}
\end{array}.} \right.
\end{equation}
\end{footnotesize}

Using the residue theorem, it can be computed :
\begin{small}
\begin{equation}
\begin{array}{*{20}{l}}
{{L^{ - 1}}\left[ {\frac{{{g_{n - i}}(s)}}{{\Delta (s)}}{e^{ - {t_0}s}}} \right]}\\
{ = \sum\limits_{j = 1}^l {{L^{ - 1}}\left[ {{e^{ - {t_0}s}}(\frac{{{c_{i,j,{d_j}}}}}{{{{(s + {s_j})}^{{d_j}}}}} +  \cdots  + \frac{{{c_{i,j,1}}}}{{s + {s_j}}})} \right]} }\\
{ = \sum\limits_{j = 1}^l {{e^{ - {s_j}(t - {t_0})}}} \left( {\frac{{{c_{i,j,{d_j}}}}}{{({d_j} - 1)!}}{{(t - {t_0})}^{{d_j} - 1}} +  \cdots  + {c_{i,j,1}}} \right)}\\
{ = \sum\limits_{j = 1}^l {{e^{ - {s_j}(t - {t_0})}}} {p_{i,j}}(t - {t_0})}
\end{array}.
\end{equation}
\end{small}

As a result, it is easy to see that 
\begin{equation}
\mathop {\lim }\limits_{t \to \infty } {L^{ - 1}}\left[ {\frac{{{g_{n - i}}(s)}}{{\Delta (s)}}{e^{ - {t_0}s}}} \right] = 0.
\end{equation}

Then we have
\begin{equation}
\begin{aligned}
& \mathop {\lim }\limits_{t \to \infty } |{z}(t) -  \bar{e}_1(t)| \\
&= \mathop {\lim }\limits_{t \to \infty } |\sum\limits_{i = 2}^{n - 1} {{L^{ - 1}}\left[ {\frac{{{g_{n - i}}(s)}}{{\Delta (s)}}{e^{ - {t_0}s}}} \right]\tilde e_i({t_0})} |   \\
& \leq \mathop {\lim }\limits_{t \to \infty } \sum\limits_{i = 2}^{n - 1}\sum\limits_{j=1}^l e^{-s_j(t-t_0)}| p_{i,j}(t-t_0)| |\tilde{e}_i(t_0) |  \\
& = 0.
\end{aligned}
\end{equation}
Obviously, $\mathop {\lim }\limits_{t \to \infty } |z(t) - {\bar e_1}(t)| = 0$ holds.  $\blacksquare$

\begin{figure}[h]
\begin{minipage}[t]{1\linewidth}
\centering
 \includegraphics[width=1\linewidth]{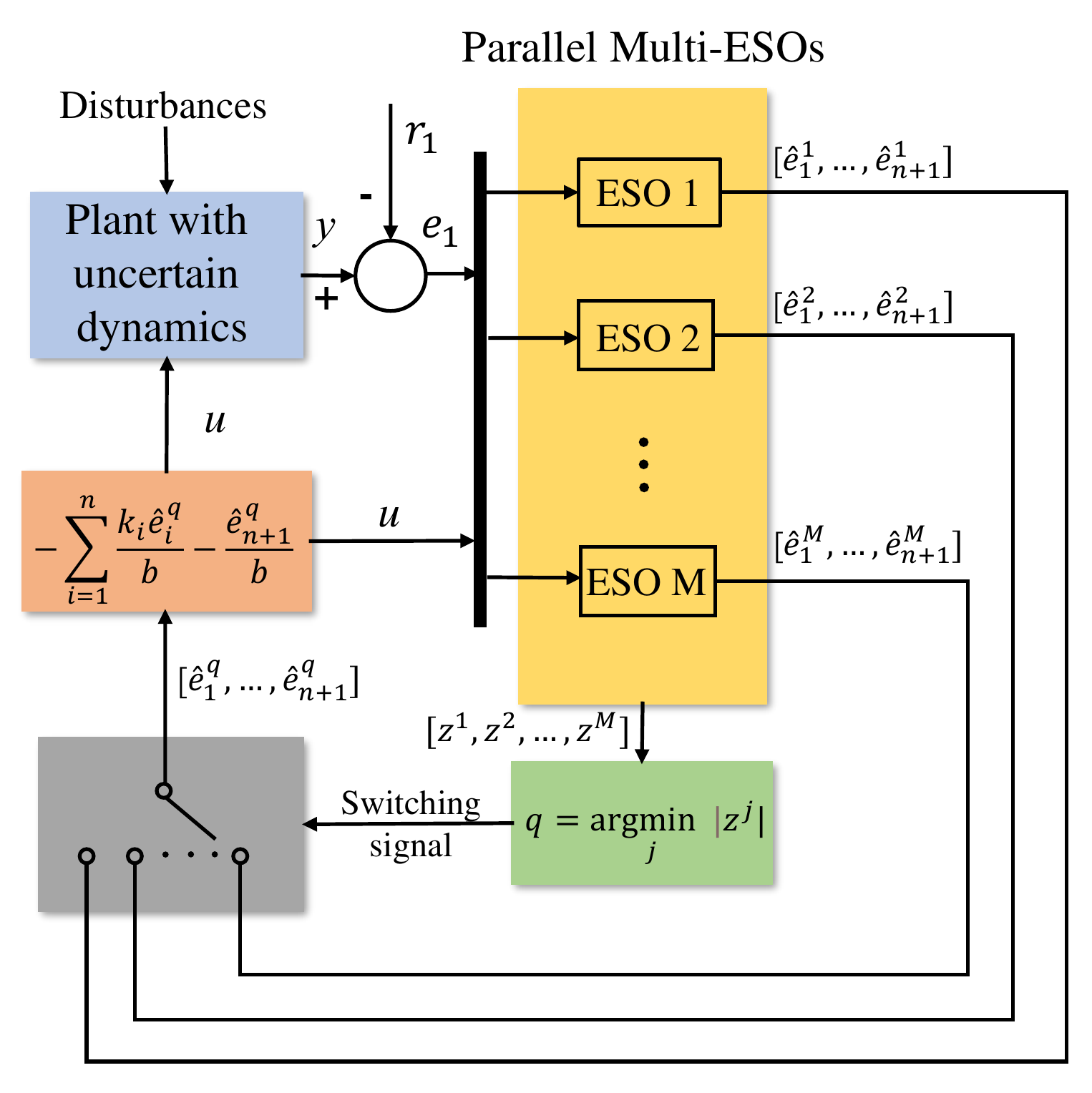}
      \caption{Control block diagram of the parallel multi-ESOs based ADRC  }
      \label{control block disgram 2}
\end{minipage}
\end{figure}
Along with the above ideas, this paper chooses $\vert \bar e_1 \vert$ as the control indicator to evaluate the tracking performance of different control laws. A subsequent problem is that we can only measure the tracking error caused by the selected control law, but how should those control laws that were not selected be evaluated? Lemma 2 and Theorem 1 give an answer: the tracking error $\bar e_1$ can be figured out by only using $\tilde e_1$. Therefore, by comparing $\vert z \vert$ caused by different control laws, we could choose the one which minimizes it.

In the following work, we propose an novel ADRC design approach, which is based on simultaneous LESOs with different orders and bandwidths. With $\vert z(t) \vert$ as the optimal indicator, the proposed algorithm is shown in Algorithm 1, in which $\hat{e}_{r}^j$ is the estimation of the $j$th LESO with respect to $e_r$. Moreover, the control block diagram is shown in Fig.~\ref{control block disgram 2}.

\begin{algorithm}\label{algorithm}
\caption{Parallel Multi-ESOs based ADRC Algorithm}
\begin{algorithmic}[t]
\STATE $\quad$
\STATE \textbf{Initialization} : Constructing $M$ LESOs with different orders and bandwidths  according to (\ref{LESO}) ;

\STATE \textbf{For every control time $t$} :
\begin{itemize}[leftmargin=*]
\setlength{\itemsep}{5pt}
\item Get the measurement $y(t) = x_1(t)$;

\item Update the values of the  LESOs and the ideal closed-loop system (\ref{ideal system});

\item Calculate $e_1^j(t)=x_1(t)-x_1^*(t)$ and $\tilde{e}_{1}^j(t)=e_1^j(t)-\hat{e}_{1}^j(t), j=1,2,...,M$;

\item Calculate $z^j(t), j=1,2,..,M$ by (\ref{approximate});

\item Select the estimation of ESO that minimizes $\vert z \vert$ by \[q = \mathop {argmin}\limits_j |{\rm{ }}z^j({t})| ; \]

\item Design control law as \[ u(t)=\dfrac{- \sum\limits_{i = 1}^n {{k_i}\hat e_i^q(t)}  - \hat e_{n + 1}^q(t)}{b}. \]
\end{itemize}

\end{algorithmic}
\end{algorithm}

\subsection{Stability Analysis for  Parallel Multi-ESOs based ADRC}

This section is going to analyse the closed-loop stability of the proposed parallel multi-LESOs based ADRC.

After ensuring the stability of the LESOs, we have the following Theorem 2 to guarantee the boundedness of tracking error.

{\bf Theorem 2.} Considering the closed-loop error system (\ref{error system}) and the parallel multi-LESOs based ADRC algorithm under the same conditions as Lemma 1, we have the following results for any $t>t_0$:
\begin{equation}\label{Theorem 3}
\begin{aligned}
\left| {{{\bar e}_1}(t)} \right| & \le {e^{{A^*}(t - {t_0})}}{_\infty }{\left\| {\bar e({t_0})} \right\|_\infty }\\
              & + \sum\limits_{j = 1}^l {\sum\limits_{i = 1}^{{d_j} - 1} {\frac{{{c_{n,j,i}}}}{{s_j^{i + 1}}}\Delta ({\omega _o})} } \gamma (t,{t_0})
\end{aligned},
\end{equation}
where $c_{n,j,i}$ is as defined in (\ref{c}), $s_j$ is as defined in (\ref{Characteristic polynomial}) and $ \gamma (t,{t_0}) = \mathop {\sup }\limits_{s \in [{t_0},t]} {\left\| {\varepsilon (s)} \right\|_\infty }.$

{\bf Remark 3.} Obviously, the first item of the right hand of (\ref{Theorem 3}) converges to 0 as $t \to \infty$. For the second item, noting that $\gamma(t,t_0)$ only 
depends on the estimation errors, it can be known that although the estimation errors of M LESOs are different, as long as their boundedness can be guaranteed, the parallel multi-ESOs based ADRC would be able to achieve the control goal under rational pole allocation.



{\bf Proof of Theorem 2.}

The explicit solution of (\ref{error system}) is
\begin{equation}
\bar e(t) = {e^{{A^*}(t - {t_0})}}\bar e({t_0}) + \int_{{t_0}}^t {{e^{{A^*}(t - \tau )}}{B}\delta (\tilde e(\tau ))} d\tau .
\end{equation}

On the one hand, 
\begin{equation}
\begin{aligned}
|\delta (\tilde e(\tau))| & \leq  {k_1}\vert {{\tilde e}_1} \vert +  \cdots  + {k_n}\vert {{\tilde e}_n} \vert + \vert {{\tilde e}_{n + 1}} \vert\\
& = {k_1}|{\varepsilon _1}| +  \cdots  + {k_n}\omega _o^{n - 1}|{\varepsilon _n}| + \omega _o^n|{\varepsilon _{n + 1}}| \\
& \leq  \Delta(\omega_o) \Vert \varepsilon (\tau ) \Vert_\infty
\end{aligned}
\end{equation}

On the other hand, it is easy to verify that
\begin{equation}
{{{\left( {sI - {A^*}} \right)}^{ - 1}} = \sum\limits_{k = 0}^{n-1} {\dfrac{{{g_{n - k-1}}(s)}}{{\Delta (s)}}{{ {{A^*}} }^{k}}} }
\end{equation}
In addition, it can be seen from matrix function theory that
\begin{equation}
{\left( {sI - A^*} \right)^{ - 1}} = \sum\limits_{k = 0}^\infty  {\dfrac{1}{{{s^{k + 1}}}}{{A^*}^k}}.
\end{equation}
Thus,
\begin{equation}
\begin{aligned}
& {L^{ - 1}}\left[ {{{\left( {sI - {A^*}} \right)}^{ - 1}}{e^{ - {t_0s}}}} \right] \\
 =& \sum\limits_{k = 0}^\infty  {\dfrac{{{{(t - {t_0})}^k}}}{{k!}}{A^*}^k} \\
   = &{e^{{A^*}(t - {t_0})}}.
\end{aligned}
\end{equation}
Then it holds
\begin{equation}
{e^{{A^*}(t - {t_0})}} = \sum\limits_{k = 0}^{n - 1} {{L^{ - 1}}\left[ {\frac{{{g_{n - k - 1}}(s)}}{{\Delta (s)}}} e^{-t_0 s} \right]} {A^*}^k. 
\end{equation} 
Therefore, we have
\begin{equation}
\begin{aligned}
& W(t) \\
\equiv & \int_{{t_0}}^t {{e^{{A^*}(t - \tau )}}{B}\delta (\tilde e(\tau ))} d\tau  \\
= & \sum\limits_{k = 0}^{n - 1} {\int_{{t_0}}^t {{L^{ - 1}}\left[ {\frac{{{g_{n - k - 1}}(s)}}{{\Delta (s)}}} e^{-\tau s} \right]{A^*}^k{B}\delta (\tilde e(\tau ))} d\tau } \\
= &  \sum\limits_{k = 0}^{n - 1} {T(k,t){N_1}(k)} \\
\end{aligned},
\end{equation}
where $N_1(k)={A^*}^kB$ and
\begin{small}
\begin{equation}
\begin{aligned}
T(k,t)  
= &  {\int_{{t_0}}^t {{L^{ - 1}}\left[ {\frac{{{g_{n - k - 1}}(s)}}{{\Delta (s)}}} e^{-\tau s} \right]\delta (\tilde e(\tau ))} d\tau } \\
= & \sum\limits_{j = 1}^l {\sum\limits_{i = 1}^{{d_j} - 1} {\int_{{t_0}}^t {{e^{ - {s_j}(t - {\tau})}}\frac{{{c_{k + 1,j,i}}}}{{i !}}{{(t - \tau)}^i}\delta (\tilde e(\tau ))d\tau } } }
\end{aligned}.
\end{equation}
\end{small}

$N_1(k)$ can be derived directly
\begin{equation}
\begin{aligned}
& N_1(0)= B = [0,0,\cdots, 0, 1]^T \\
& N_1(1)= A^*B = [0,0,\cdots,1, -k_n]^T, \\
& \vdots \\
& N_1(n-1) = {A^*}^{n-1}B = [1, -k_n,\cdots,*,*]^T,
\end{aligned}
\end{equation}
where ``*'' are unknown items we do not care. Then it can be obtained
\begin{equation}
W(t) = \left[ {\begin{array}{*{20}{c}}
  {T(n - 1,t)} \\ 
  * \\ 
   \vdots  \\ 
  * 
\end{array}} \right].
\end{equation}
Let ${e^{{A^*}(t - {t_0})}}\bar e({t_0}) = {\left[ {{q_1}(t),{q_2}(t), \cdots ,{q_n}(t)} \right]^T},$ and then
\begin{equation}
{\bar e_1}(t) = {q_1}(t) + T(n - 1,t).
\end{equation}

Now we are going to study the properties of $T(n-1,t)$. Calculating directly, we have
\begin{equation}
\begin{array}{l}
\int_{{t_0}}^t {{e^{ - {s_j}(t - \tau )}}{{(t - \tau )}^i}} d\tau \\
 = \frac{1}{{s_j^{i + 1}}}\left[ {i! - {e^{ - {s_j}(t - {t_0})}}\left( {\sum\limits_{k = 1}^{i + 1} {\frac{{i!}}{{(i + 1 - k)!}}{{\left( {{s_j}(t - {t_0})} \right)}^{i + 1 - k}}} } \right)} \right].
\end{array}
\end{equation}

Obviously, the right hand of the equation is monotonically increasing to $\dfrac{i!}{s_j^{i+1}}$ on $[t_0, \infty)$. Thus,
\begin{equation}
\left| {T(n - 1,t)} \right| \leq \sum\limits_{j = 1}^l {\sum\limits_{i = 1}^{{d_j} - 1} {\frac{{{c_{n,j,i}}}}{{s_j^{i + 1}}}\Delta ({\omega _o})} } \gamma (t,{t_0}),
\end{equation}
where 
\begin{equation}
\gamma (t,{t_0}) = \mathop {\sup }\limits_{s \in [{t_0},t]} {\left\| {\varepsilon (s)} \right\|_\infty }.
\end{equation}

Finally,
\begin{equation}
\begin{aligned}
\left| {{{\bar e}_1}(t)} \right| & \le {e^{{A^*}(t - {t_0})}}{_\infty }{\left\| {\bar e({t_0})} \right\|_\infty }\\
              & + \sum\limits_{j = 1}^l {\sum\limits_{i = 1}^{{d_j} - 1} {\frac{{{c_{n,j,i}}}}{{s_j^{i + 1}}}\Delta ({\omega _o})} } \gamma (t,{t_0})
\end{aligned}. \blacksquare
\end{equation}


\section{Application to High-Speed Precision Motion Stage}
 As the core module of intelligent manufacturing, high speed precision motion stage (HSPMS) is widely used in integrated circuit manufacturing, microelectronics processing technology, semiconductor processing industry \cite{inproceedings6,inproceedings7, article25}. Due to the high-stiffness structure, traditional high-speed precision motion stage (HSPMS) is  affected by high frequency disturbance near the dead zone of friction. The friction is a very complex non-linear phenomenon which can cause large steady-state error and oscillation. In order to overcome this shortcoming, \cite{article15} designed the rigid-flexible coupling (RFC) stage, which can convert high-frequency friction disturbance into low-frequency friction disturbance through special structural design. Moreover, it proved that the traditional LESO based ADRC can estimate and compensate low frequency disturbances effectively. In the following researches, the parallel multi-ESOs based ADRC will be verified in the RFC stage.

\subsection{Experimental Setup and Dynamical Model}
\begin{figure}[h]\centering
\begin{minipage}[t]{0.9\linewidth}
\centering
 \includegraphics[width=1\linewidth]{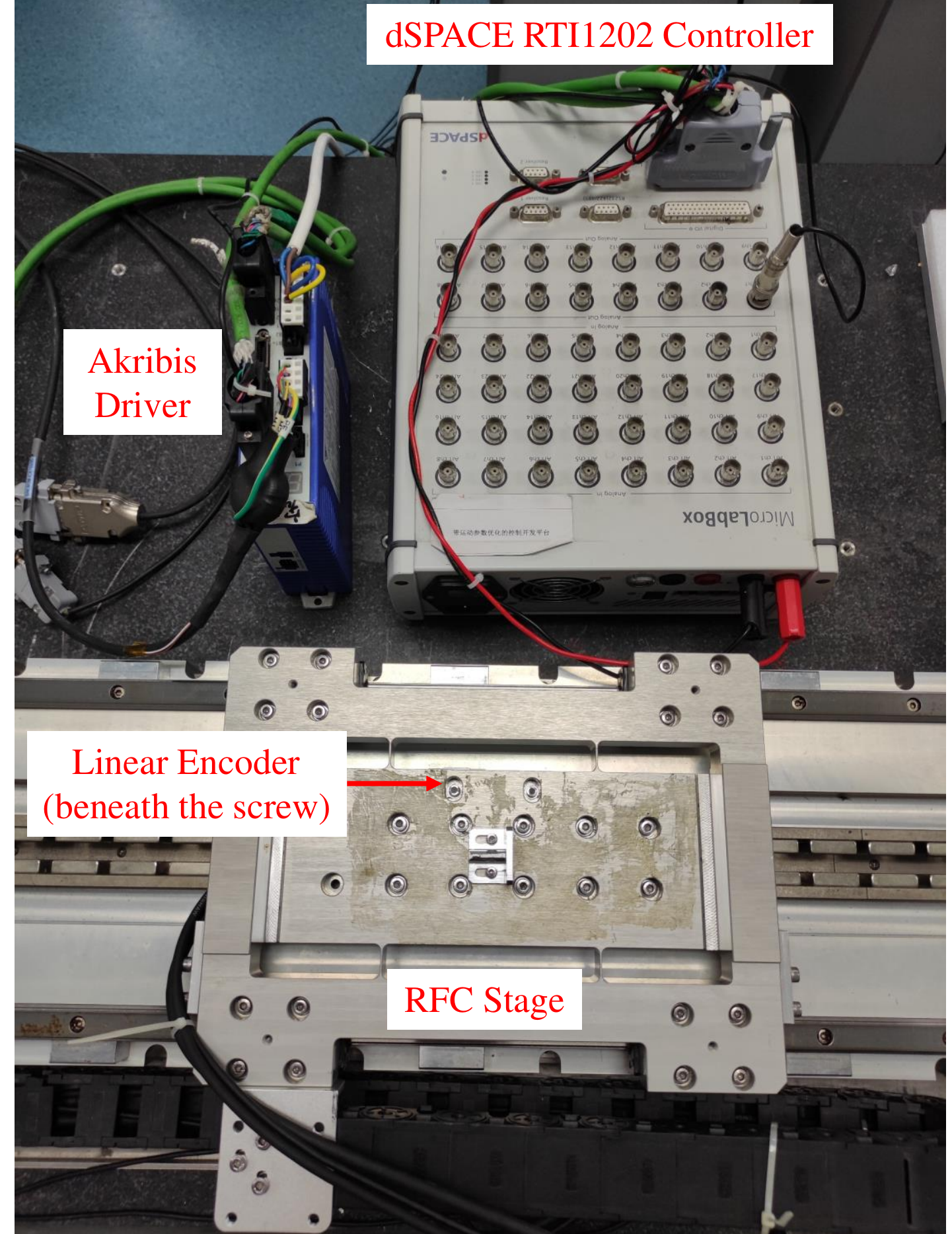}
      \caption{Experimental equipment  }
      \label{RFC1}
\end{minipage}
\end{figure}
The chosen experimental equipment is the RFC stage motion control system shown in Fig.~\ref{RFC1}, which consists of dSPACE RTI1202 controller, Akribs ASD driver, Renishw incremental linear encoder (0.1 $\mu$m resolution) and the RFC stage. The workflow of this equipment is shown in Figure 1.
\begin{figure}[h]
\begin{minipage}[t]{1\linewidth}
\centering
 \includegraphics[width=1\linewidth]{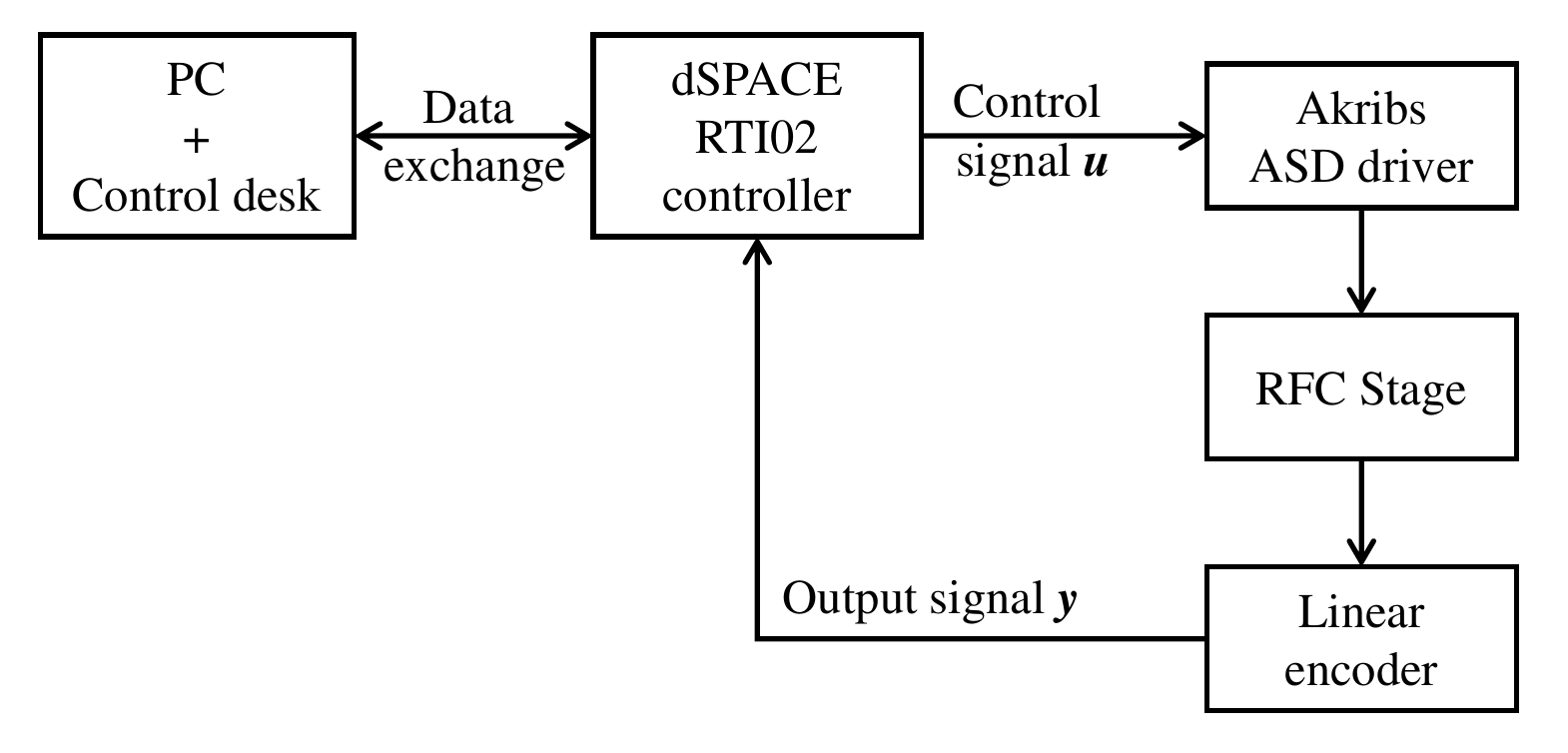}
      \caption{Workflow of the experimental equipment   }
      \label{Control loop}
\end{minipage}
\end{figure}

\begin{figure}[h]
\begin{minipage}[t]{1\linewidth}
\centering
 \includegraphics[width=1\linewidth]{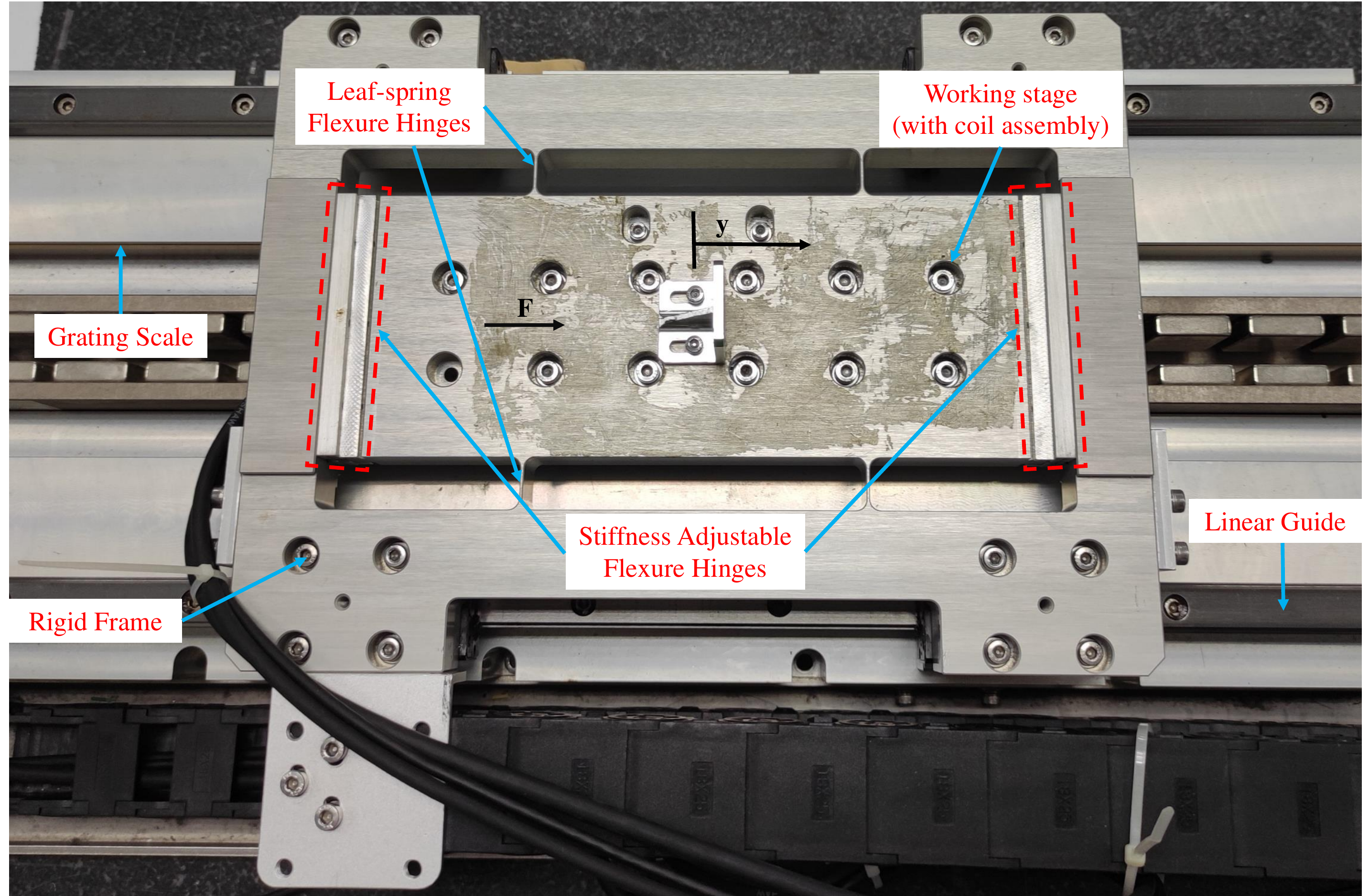}
      \caption{Experimental devices  }
      \label{RFC2}
\end{minipage}
\end{figure}

As shown in Fig.~\ref{RFC2}, the RFC stage consists of two parts: a working stage and a rigid frame, which are connected by flexible hinges. When the rigid frame is in the friction dead zone, the displacement of the working stage is realized by the elastic deformation of the flexure hinge. The working stage, the leaf spring flexure hinge and the rigid frame are machined in one piece, made from aero duralumin 7075 to eliminate assembly errors. The flexible hinges with adjustable stiffness are made of 65mn spring steel.  After assembly, the table and rigid frame are connected together from both ends to ensure fatigue life. Besides that, the linear encoder and coil assembly are installed on the working stage to measure the displacement output $y$ and receive the control input $u$. The rigid frame is installed on the linear guide rail to achieve large stroke movement.

\begin{figure}[h]
\begin{minipage}[t]{1\linewidth}
\centering
 \includegraphics[width=1\linewidth]{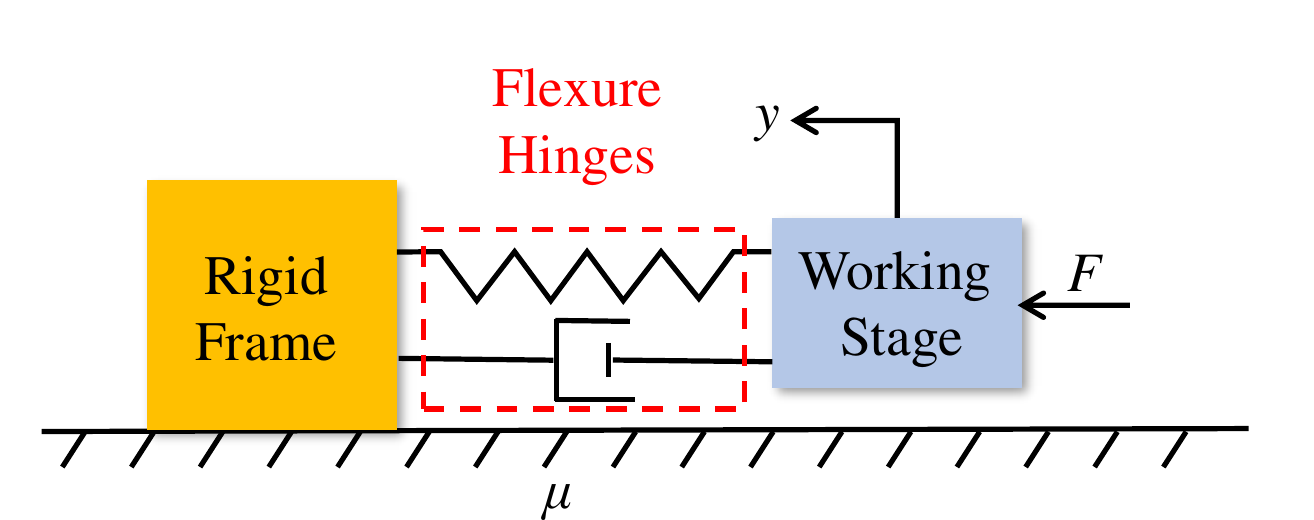}
      \caption{Equivalent mechanics model of RFC positioning stage  }
      \label{equivalent mechanics}
\end{minipage}
\end{figure}

The equivalent mechanics model of the RFC positioning stage is shown in Fig.~\ref{equivalent mechanics}. We can see that the flexure hinges keep the working stage free from friction and only rigid frame is affected by friction.

Since the input and output of the system are all on the working stage, the following study only takes the working stage as the plant for simplicity. All
the effects caused by the rigid frame are applied to the working stage as disturbance of flexure hinge. Hence, the dynamic model of the RFC positioning stage can be described as
\begin{equation}\label{RFC model}
\begin{cases}
\dot x_1=x_2 \\
\dot x_2=\dfrac{k_a k_s}{m}u + \dfrac{kd_x+cd_v}{m} + f(x_1,x_2,t) \\
y=x_1 
\end{cases}.
\end{equation}
The physical meanings of each symbol are shown in TABLE I.

\begin{table}[H]\centering
\caption{ Physical meanings of each symbol}
\normalsize
\begin{tabular}{|c| c|}
\hline
Symbol & Physical Meaning \\
\hline
$x_1$       & Displacement of the working stage          \\
\hline
$x_2$  &  Velocity of the working stage   \\
\hline
$k_a$  &  Proportion from voltage to current     \\
\hline
$k_s$  &  Force constant of linear motor     \\
\hline
$m$  &  Mass of the working stage     \\
\hline
$u$  &  Analog voltage     \\
\hline
$k$  & Equivalent stiffness of flexure hinges   \\
\hline
$c$  & Equivalent damping of flexure hinges     \\
\hline
$d_x$  & Flexure hinges' elastic deformation   \\
\hline
$d_v$  & Flexure hinges' elastic deformation rate  \\
\hline
$f$  & Uncertainty and disturbance  \\
\hline
$y$  & Measurement output  \\
\hline
\end{tabular}
\end{table}

\subsection{Experiment of Point-to-Point Motion}
In the field of electronics manufacturing and laser processing industry, high precision is the biggest characteristic of point-to-point motion. Therefore, in the following experiment, we consider the displacement control of the working stage, that is, let $x_1$ track the command signal $r$.

The ideal closed-loop system is set as (\ref{ideal system}) with {$\omega_c=150$, $k_1={\omega_c^2}$ and $k_2=2\omega_c$. Taking $e_3=f+\frac{kd_x+cd_v}{m}-\ddot{r}$ as the ``total disturbance'', the following 3rd-order LESO (\ref{3}) and 4th-order LESO (\ref{4}) with $\omega_o = 1500$  are set to estimate $e_1 = x_1 - r$, $e_2=\dot{e}_1$ and $e_3$. Besides that, the input gain $b = \frac{k_a k_s}{m}$ is 3.25.

\begin{equation}\label{3}
\begin{cases}
\dot{\hat e}_{11} = \hat e_{12} + 3\omega_o \left( e_1 - \hat{e}_{11} \right) \\
\dot{\hat e}_{12} = \hat{e}_{13} + 3{\omega_o}^2\left( e_1 - \hat{e}_{11} \right) + bu \\
\dot{\hat e}_{13}= {\omega_o}^3\left( e_1 - \hat{e}_{11} \right) 
\end{cases},
\end{equation}

\begin{equation}\;\label{4}
\begin{cases}
\dot{\hat e}_{21} = \hat e_{22} + {4\omega_o} \left( e_1 - \hat{e}_{21} \right) \\
\dot{\hat e}_{22} = \hat{e}_{23} + 6{\omega_o}^2\left( e_1 - \hat{e}_{21} \right) + bu \\
\dot{\hat e}_{23}= \hat{e}_{24}+ 6{\omega_o}^3\left( e_1 - \hat{e}_{21} \right) \\
\dot{\hat e}_{24}= {\omega_o}^4\left( e_1 - \hat{e}_{21} \right) \\
\end{cases}.
\end{equation}

Correspondingly, the approximate formula (\ref{approximate}) can be written as
\begin{equation}
{z^j}(s) = \frac{{{s^2} + ({\beta _{j1}} + {k_2})s + {\beta _{j2}} + {k_2}{\beta _{j1}} + {k_1}}}{{{s^2} + {k_2}s + {k_1}}}{\tilde e_{j1}}(s),
\end{equation}
for $j=1, 2$.
However, due to the influence of measurement noise, there may be errors in the calculation of $z$, which may cause frequent switches between the two ESOs and result in frequent and drastic changes in the input. To overcome this problem, in the actual experiment, the switching judgment is executed every 20 sampling periods and the switching index is changed to the cumulative value of $\vert z \vert$ accordingly.

To verify the proposed parallel multi-ESOs based ADRC method, the above LESOs are run at the same time. In addition, the respective traditional ADRC based on these 2 LESOs are set as comparisons.

\subsection{Experimental Results}

Fig.~\ref{error1} shows the tracking errors of different control laws when $r=10$ is taken as the reference signal. Compared with the ADRC based on 3rd-order LESO, the proposed method has a smaller tracking error in the initial stage and enters the steady state faster.  Compared with the ADRC based on 4th-order LESO, the proposed method has a smoother error curve, which means that the system output is smoother. Besides that, the IAE of the above three methods are shown in Table II.

\begin{figure}[h]
\begin{minipage}[t]{1\linewidth}
\centering
 \includegraphics[width=1\linewidth]{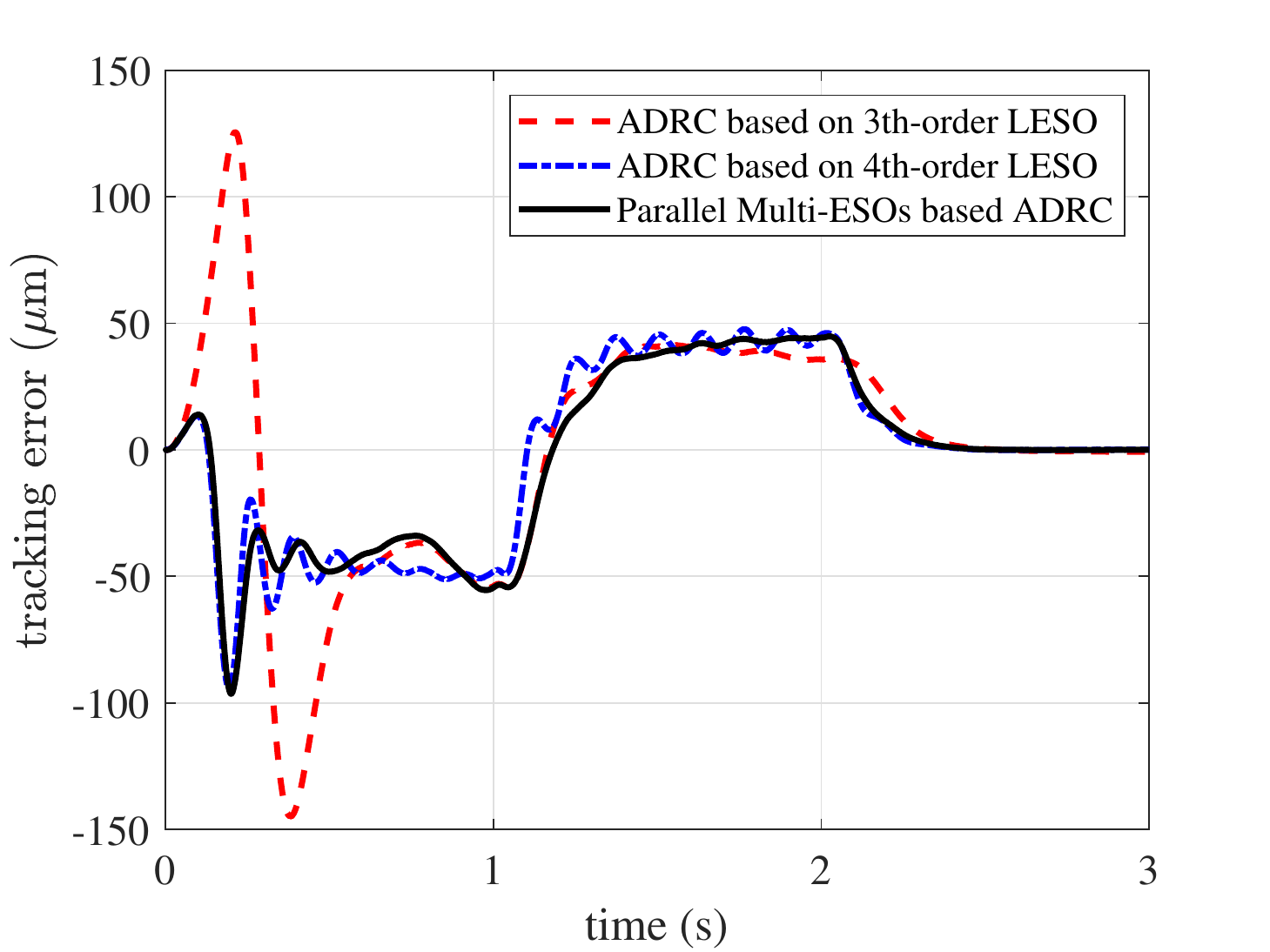}
      \caption{Curves of tracking error under different control laws }
      \label{error1}
\end{minipage}
\end{figure}

\begin{table}[h]\centering
\caption{IAE of different control laws}
\normalsize
\begin{tabular}{|c|c|}
\hline
Control law       &     IAE    \\
\hline
 ADRC based on 3rd-order LESO & 105.95 \\
\hline
ADRC based on 4th-order LESO &  86.15 \\
\hline
Parallel Multi-ESOs based ADRC & 82.67 \\
\hline
\end{tabular}
\end{table}

Fig.~\ref{u} shows the curves of input and switching signal,  where switching signal ``0'' represents the 3rd-order LESO and ``1'' represents the 4th-order LESO. As can be seen from the figure, when the control law switches between the two LESOs, there will be a small fluctuation in the system input. The maximum fluctuation occurs at t=0.2s and the voltage changes from -1.1V to 0.6V. Considering that the set voltage variation range is -10$\sim$10V, the variation of 1.7V is acceptable. Moreover, combined with Fig.~\ref{error1} and Fig.~\ref{u}, it can be seen that the sudden change of input has no significant influence on the closed-loop system.

\begin{figure}[h]
\begin{minipage}[t]{1\linewidth}
\centering
 \includegraphics[width=1\linewidth]{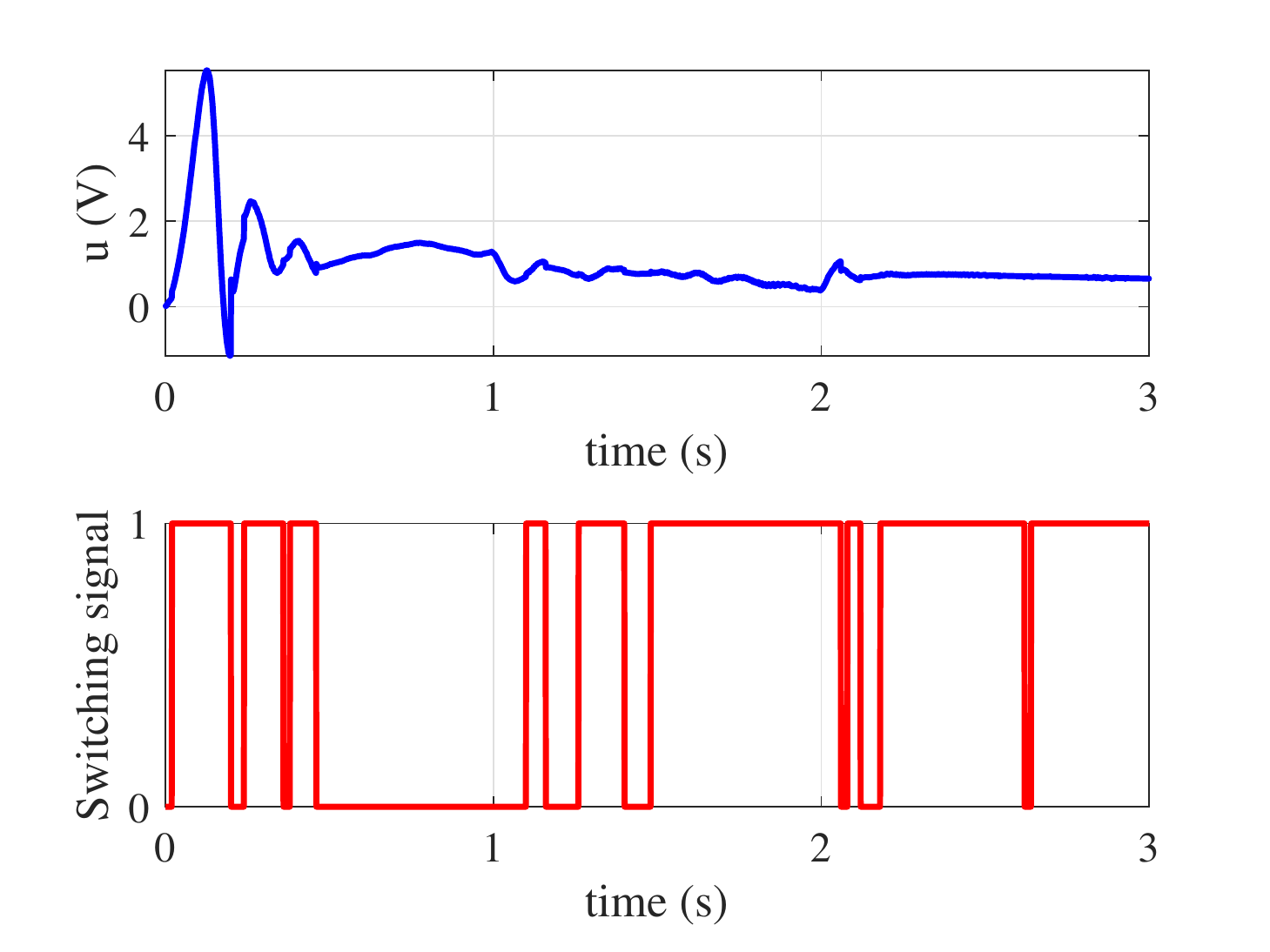}
      \caption{Curves of input and switching singal }
      \label{u}
\end{minipage}
\end{figure}

Considering the difference of friction disturbance in different parts of mechanical bearing and the randomness near the friction dead zone, it is necessary to make repeated experiment with different trajectories. Fig.~\ref{error2} demonstrates the tracking error curves under $r=10$ and $r=20$, respectively, each repeated 5 times. On the one hand, the dynamic characteristics of tracing error curves are very similar for different reference signals. On the other hand, the error curves are almost the same for 5 repeated experiments for the same reference signal. All these results show the robustness of the parallel multi-ESOs based ADRC. Furthermore, we also conducted 5 repeated experiments on the two control groups, and their mean IAEs are shown in Table III.

\begin{figure}[h]
\begin{minipage}[t]{1\linewidth}
\centering
 \includegraphics[width=1\linewidth]{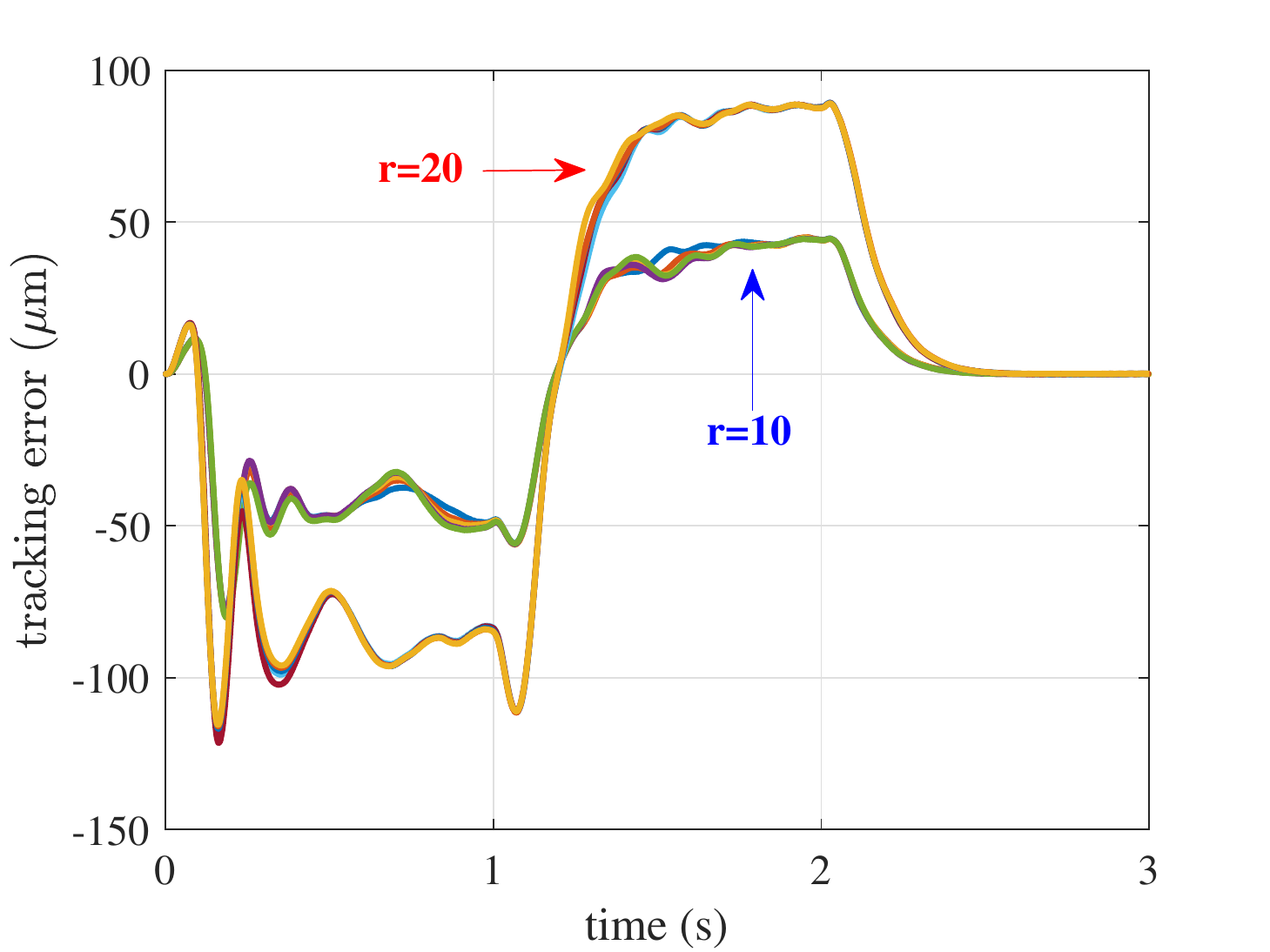}
      \caption{Repeated experiments with different trajectories of RFC positioning stage }
      \label{error2}
\end{minipage}
\end{figure}

\begin{table}[h]\centering
\caption{Mean IAE of different methods}
\normalsize
\begin{tabular}{|c|c|c|}
\hline
 \diagbox{Control law}{Mean IAE}{Reference signal}               & $r=10$      &     $r=20$    \\ 
\hline
 ADRC based on 3rd-order LESO & 101.29 &  176.46 \\
\hline
ADRC based on 4th-order LESO  &  86.00 & 170.34 \\
\hline
Parallel Multi-ESOs based ADRC & 82.18 & 164.78 \\
\hline
\end{tabular}
\end{table}

\section{Conclusion}
From the obtained results, one can see that the estimation error of ESO and the tracking error of ADRC are not exactly positively correlated, which suggests that in order to improve the control performance of ADRC, it is not enough to only consider optimization of ESO. Therefore, this paper directly starts from the control effect, by setting up parallel ESOs to switch different control laws online, so as to achieve an improvement on the traditional ADRC. The stability and superiority of the method are proved by theory and experimental results.

\section*{Acknowledgments}
This work is supported by the National Natural Science Foundation of China under Grant U20A6004, 51875108 and National Key R\&D Program of China
( 2022YFB4701001).

\bibliographystyle{IEEEtran}
\bibliography{ref}

\end{document}